\documentclass[amstex,a4]{svmult}

\usepackage{amssymb}
\usepackage{graphicx}
\usepackage{graphics}
\usepackage{epsfig}
\usepackage{psfrag}
\usepackage{times}
\usepackage{bbm}

\begin{document}

\newcommand{\be}{\begin{eqnarray}}
\newcommand{\ee}{\end{eqnarray}}
\newcommand{\bea}{\begin{eqnarray}}
\newcommand{\eea}{\end{eqnarray}}
\newcommand{\bma}{\begin{subequations}}
\newcommand{\ema}{\end{subequations}}
\def\qp{\pm}
\def\qed{\leavevmode\unskip\penalty9999 \hbox{}\nobreak\hfill
     \quad\hbox{\leavevmode  \hbox to.77778em{%
               \hfil\vrule   \vbox to.675em%
               {\hrule width.6em\vfil\hrule}\vrule\hfil}}
     \par\vskip3pt}
\def\lR{l^2_{\mathbb{R}}}
\def\RR{\mathbb{R}}
\def\E{\mathbf e}
\def\D{\boldsymbol \delta}
\def\S{{\cal S}}
\def\T{{\cal T}}
\def\dd{\delta}
\def\E{{\bf E}}
\newcommand{\Q}[1]{{{\footnotesize{\emph{[ #1 ] }}}}}
\newcommand{\M}[1]{{\marginpar{\footnotesize{\emph{ #1  }}}}}
\newcommand{\tr}[1]{{\rm tr}\left[#1\right]}
\newcommand{\fett}[1]{\ {\\{\bf{#1}}}}
\newcommand{\C}{{\Bbb C}}

\def\id{\mathbbm 1}

\chapter{Quantum Computing}

\vspace*{-0.7cm}
\noindent
{\small
J.\ Eisert$^{1,2}$ and M.M.\ Wolf$^{3}$\\

\smallskip
\noindent
1 Universit{\"a}t Potsdam, Am Neuen Palais 
10, 14469 Potsdam, Germany\\
\noindent
2 Imperial College London, Prince Consort Road, 
SW7 2BW London, UK\\
\noindent
3 Max-Planck-Institut f{\"u}r Quantenoptik, 
Hans-Kopfermann-Str.\ 1, 85748 Garching, Germany\\
}

\smallskip

Quantum mechanics is one of the cornerstones of modern 
physics. It 
governs the behavior and the properties of matter in a fundamental
way, in particular on the microscopic scale of atoms and
molecules. Hence, what we may call a classical computer,
i.e., one of 
those machines on or under the desktops in our offices
together with all their potential descendants, is itself
following the rules of quantum mechanics. However, 
such devices are no
quantum computers in the sense that all the inside
information processing  can perfectly be described 
within classical information theory. In fact, we do not need
quantum mechanics in order to explain how the zeros and ones -- the
bits -- inside a classical computer evolve. The reason for
this is that the architecture of classical computers does not make
use of one of the most fundamental features of quantum mechanics,
namely the possibility of superpositions. Throughout the
entire processing of any program on a classical computer, each of
the involved bits takes on either the value zero or one. Quantum
mechanics, however, would in addition allow superpositions of
zeros one ones, that is, bits -- now called qubits
(quantum-bits) -- which are somehow in the state zero and one at
the same time. Computing devices which exploit this possibility,
and with that all the essential features of quantum mechanics, are
called quantum computers \cite{Mike}.
Since they have an additional
capability they are at least as powerful as classical computers: every
 problem that can be solved on a classical computer can be handled
 by a quantum computer just as well. The converse, however, is
 also true since the dynamics of quantum systems is governed by
 linear differential equations, which can in turn be solved (at least
 approximately) on a classical computer. Hence, classical and
 quantum computers could in principle emulate each other and
   quantum computers are thus no
 hypercomputers\footnote{A \emph{hypercomputer} would be
capable of solving problems that can not be handled by a
\emph{universal Turing machine}
 (the paradigm of a classical digital computer). The most famous
example of such a problem is the \emph{halting problem} which is in
modern terminology the task of a universal crash debugger, which is
supposed to spot all bugs leading to crashes or infinite loops for
any program running on a universal Turing machine.
As shown by Turing such a debugger cannot exist.}.
So why quantum computing? And if there is any reason, why
not just simulate these devices (which do not exist yet anyhow) on
a classical computer?

\section{Why quantum computing?}

\subsection{Quantum computers reduce the complexity of certain
computational tasks} One reason for aiming at building 
quantum computers is that they
will solve certain types of problems faster than any (present or
future) classical computer -- it seems that the border between
\emph{easy} and \emph{hard} problems is different for quantum
computers than it is for their classical counterparts. Here easy
means that the time for solving the problem grows polynomially
with the length of the input data (like for the problem of
multiplying two numbers), whereas hard problems are those for
which the required time grows exponentially. Prominent examples
for hard problems are the travelling salesman problem, the graph
isomorphism problem, and the problem of factoring a number into
primes\footnote{These problems are strongly believed to be hard
(the same is by the way true for a special instance of the
computer game ``minesweeper''). However, in all cases there is no
proof that a polynomial-time algorithm can not exist. The question
whether there exists such an algorithm (for the travelling
salesman or the minesweeper problem) is in fact the notorious
$P\stackrel{?}{=}NP$ question for whose solution there is even a
prize of one million dollar.}. For the latter it was, to the
surprise of all, shown by Peter Shor in 1994 that it could
efficiently be solved by a quantum computer in polynomial time
\cite{Shor}.
Hence, a problem which is hard for any classical computer becomes
easy for quantum computers\footnote{In fact Shor's algorithm
strikes the \emph{strong Church-Turing thesis}, which states that
every reasonable physical computing device can be simulated on a
probabilistic Turing machine with at most polynomial overhead.}.
Shor's result gets even more brisance from the fact that the
security of public key encryption, i.e., the security of home
banking and any other information transfer via the internet, is
heavily based on the fact that factoring is a hard problem.

One might think that the cost for the gained exponential speedup
in quantum computers is an exponential increase of the required
accuracy for all the involved operations. This would then be
reminiscent of the drawback of analogue computers. Fortunately,
this is not the case and a constant accuracy is sufficient.
However, achieving this ``constant'' is without doubt
experimentally highly challenging.

\subsection{Quantum systems can efficiently simulate other quantum
systems}

Nature provides many fascinating collective quantum
phenomena like superconductivity, magnetism and Bose-Einstein
condensation. Although all properties of matter are described 
by and
can in principle be determined from the laws of quantum mechanics,
physicists have very often serious difficulties to understand them
in detail and to predict them by starting from fundamental rules
and first principles. One reason for these difficulties is the
fact that the number of parameters needed to describe a
many-particle quantum system grows exponentially with the number
of particles. Hence, comparing a theoretical model for the
behavior of more than, say, thirty particles with experimental
reality is not possible by simulating the theoretical model
numerically on a classical computer without making serious
simplifications.

When thinking about this problem of simulating quantum systems on
classical computers  Richard Feynman came in the early eighties to
the conclusion that such a classical simulation typically suffers
from an exponential slowdown, whereas another quantum system could
in principle do the simulation efficiently with bearable overhead
\cite{Feynman}.

In this way a quantum computer operated as a \emph{quantum
simulator} could be used as a link between theoretical models
which are formulated on a fundamental level and
 experimental observations. Similar to Shor's algorithm a quantum
 simulator would yield an exponential speedup compared to a
 classical computer. An important difference between these two
 applications is, however, that a useful Shor-algorithm quantum
computer
 requires thousands of qubits whereas a few tens of qubits
 could already be  useful for the simulation of quantum systems. We
 will resume the idea of a quantum simulator in sections
\ref{SecAQC}, \ref{SecQS}.

 \subsection{Moore's law has physical limits} Apart from the
 computational power of a quantum computer there is a much more
 banal argument for incorporating quantum mechanics into computer
 science: \emph{Moore's law}. In 1965 Intel co-founder Gordon
 Moore observed an exponential growth in the number of transistors
 per square inch on integrated circuit and he predicted that this
trend would
 continue \cite{Moore}.
 In fact, since then this density has doubled approximately
every 18
 months\footnote{Actually, not every prediction of the pioneers in
computer business was that foresighted:
 In 1943 Thomas Watson, chairman of IBM, for instance predicted a
world market for five computers and in 1977 Digital Equipment Corp.
founder
 Ken Olson stated that ``there is no reason anyone would want a
computer in their home''. }. If this trend continues then around the
year 2020 the components
 of computers are at the atomic scale where quantum effects are
  dominant. We have thus to inevitably  cope with these effects,
  and we can either try to circumvent and eliminate them as long
  as this is possible and keep on doing classical computing or we
  can at some point try to make use of them and start doing quantum
  computing.

\subsection{Even small quantum circuits may be useful} Besides the
quantum computer with its mentioned applications quantum
information science yields a couple of other useful applications
which might be easier to realize. The best example is quantum
cryptography which enables one to transmit information with ``the
security of nature's laws'' \cite{Crypto}. However, small building
blocks of a quantum computer, i.e., small quantum circuits may be
useful as well. One potential application is for instance in
precision measurements like in atomic clocks
\cite{Precision2,Precision}. The latter are important in global
positioning systems as well as in synchronizing networks and
distant telescopes. By generating quantum correlations between the
$N$ relevant atoms in the atomic clock, a quantum circuit could in
principle reduce the uncertainty of the clock by a factor
$\sqrt{N}$.

Another application of small quantum circuits is
\emph{entanglement distillation}: in order to distribute entangled
states over large distances we have to send them through
inevitably noisy channels, thereby loosing some of the
entanglement. Fortunately, however, we can in many cases
\emph{distill} a few highly entangled states out of many weakly
entangled ones \cite{Distill, Distill2}.

\section{From classical to quantum computing}\label{QvsC}

Let us now have a closer look at the way a quantum computer works.
We will do so by comparing the concepts of classical computing
with the basics of quantum computing. In fact, many classical
concepts have very similar quantum counterparts, like bits become
qubits and still the logic is often best explained within a
circuit model \cite{Deutsch, Mike}. However, there are also
crucial differences, which we will describe on the following
pages.

\subsection{Qubits and quantum parallelism} The elementary information
carriers in a quantum computer are the \emph{qubits} -- quantum
bits \cite{Qubit}. In contrast to classical bits which take on
either the value zero or one, qubits can be in every
\emph{superposition} of the state vectors $|0\rangle$ and
$|1\rangle$. This means that the vector $|\Psi\rangle$ describing
the (pure) state of the qubit can be any linear combination
\begin{equation}|\Psi\rangle\ =\ \alpha|0\rangle +
\beta|1\rangle\label{Psi}\end{equation} of the vectors $|0\rangle$
and $|1\rangle$ with complex coefficients $\alpha$ and
$\beta$.\footnote{The ``Dirac notation'' $|\cdot\rangle$ is
frequently used in quantum mechanics. Eq.\ \ (\ref{Psi}) could as well
be written in the standard vector notation, i.e., $\Psi=(\alpha,
\beta)$ such that $|0\rangle$ and $|1\rangle$ correspond to the
basis vectors $(1,0)$ and $(0,1)$ respectively.}
In the same way a system of many qubits can be in a superposition
of \emph{all} classically possible states
\begin{equation}\label{super}
|0,0,\ldots,0\rangle + |1,0,\ldots,0\rangle + \ldots +
|1,1,\ldots,
1\rangle\;.
\end{equation}
The basis $\{|0,0, ...,0\rangle,|0,1,...,0\rangle,...,|1,1,
...,1\rangle\}$ that corresponds to the binary words of length $n$
in a quantum system of $n$ qubits is called the {\em computational
basis}\footnote{In finite-dimensional quantum systems as we encounter here
the computational basis spans the {\em Hilbert space} associated with
the physical system.}.

Using the superposition of Eq.\ (\ref{super}) as an input
for an algorithm means somehow to run the computation on all
classically possible input states at the same time. This
possibility is called \emph{quantum parallelism} and it is
certainly one of the reasons for the computational power of a
quantum computer. The mathematical structure behind the
composition of quantum systems is the one of the tensor product.
Hence, vectors like $ |0,0,\ldots,0\rangle$ should be understood
as $|0\rangle \otimes ...\otimes |0\rangle=|0\rangle^{\otimes n}$.
This implies that the dimension of the space characterizing the
system grows exponentially with the number of qubits. A

Physically, qubits correspond to effective two-level systems like
 ground state and excited state of an atom, the polarization
 degree of freedom of light or up-and down orientation of a
 spin-1/2 particle, see section \ref{Secimp}.
 Such a physical system
 can be in any \emph{pure state} that can be represented by a
 normalized
  vector of the above form\footnote{States in quantum mechanics, however,
  can also be \emph{mixed}, in contrast to pure states which can
 be represented as state vectors.
 A general and hence mixed quantum
 state can be represented by a
 \emph{density operator} $\rho$. A density operator $\rho$
is a positive operator, $\rho\geq 0$,
 which  is normalized, ${\rm tr}[\rho]=1$. For qubits, the state space, i.e.,
the set of all possible density matrices representing possible
physical states, can be represented as a unit ball, called the
\emph{Bloch ball}. The extreme points of this set are the pure
states that correspond to state vectors. In the Bloch picture, the
pure states are located on the boundary of the set: the set of all
pure states is hence represented by a unit sphere. The concept of
mixed quantum states is required in quantum mechanics to
incorporate classical ignorance about the preparation procedure,
or when states of parts of a composite quantum system are
considered. }. A pure state of a composite quantum system that is 
not a product with respect to all constituents is called an
\emph{entangled} pure state.

 \subsection{Read out and probabilistic nature of quantum
computers}
An important difference between classical and quantum computers
lies
in the read-out process. In the classical case there is not much
to say: the output is a bit-string which is obtained in a
deterministic manner, i.e., repeating the computation will lead to
the same output again\footnote{Within the circuit model described
above this is a trivial observation since all the elementary gates
are deterministic operations. Note that even \emph{probabilistic}
classical algorithms run essentially on deterministic grounds.}.
However, due to the probabilistic nature of quantum mechanics,
this is different for a quantum computer. If the output of the
computation is for instance the state vector
$|\Psi\rangle$ in
Eq.\  (\ref{Psi}), $\alpha$ and $\beta$ cannot be determined by a
single measurement on a single specimen.
In fact, $|\alpha|^2$ and $|\beta|^2$ are the
probabilities for the system to be found in $|0\rangle$
and $|1\rangle$ respectively. Hence, the absolute values of these
coefficients can be determined by repeating the computation,
measuring in the basis $|0\rangle,\;|1\rangle$ and then counting
the relative frequencies. The actual outcome of every single measurement is thereby completely indetermined.
In the same manner, the state of a quantum system
consisting of $n$ qubits can be measured in the
computational basis, which means that
the outcome corresponding to some binary word occurs
with the probability given by the square of the absolut
value of the respective coefficient.
So in effect, the probabilistic nature of the read out process
on the one hand and the possibility of exploiting quantum
parallelism on the other hand are competing aspects when
it comes to comparing the computational power of quantum
and classical computers.

\subsection{The circuit model}
A classical digital computer operates on a string of input bits
and returns a string of output bits. The function in between can
be described as a logical circuit build up out of many elementary
logic operations. That is, the whole computation can be decomposed
into an array of smaller operations -- gates -- acting only on one
or two bits like the AND, OR and NOT operation. In fact, these
three gates together with the COPY (or FANOUT) operation form a
\emph{universal} set of gates into which every well-defined
input-output function can be decomposed.
The complexity of an algorithm is then essentially the number of
required elementary gates, resp. its asymptotic growth with the
size of the input.

\begin{figure}
\centerline{
\includegraphics[width=7cm]{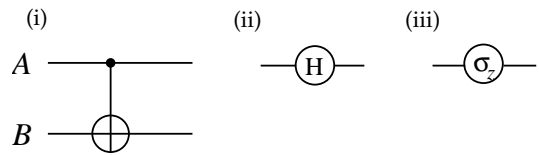}
}
    \caption{Representation of a (i) quantum CNOT gate, a (ii) Hadamard
gate, and (iii) a Pauli $\sigma_z$ gate.
    In the CNOT gate the first qubit, here denoted as $A$, is typically
referred
    to as control, the second qubit $B$ as target. The CNOT gate is a
quantum version of the XOR gate, made
    reversible by   retaining the control.}\label{fig2}
\end{figure}

The circuit model for the quantum computer \cite{Deutsch, Mike} is
actually very reminiscent of the classical circuit model: of
course, we have to replace the input-output function by a quantum
operation mapping quantum states onto quantum states. It is
sufficient to consider operations only that have the property to
be unitary, which means that the computation is taken to be
logically reversible.
In turn, any unitary operation can
be decomposed
into elementary gates acting only on one or
two qubits. A set of elementary gates that allows
for a realization of any unitary to arbitrary approximation is
again referred to as being {\em universal} \cite{Gates,Deutsch}.
An important example of a set of universal gates is in
this case any randomly chosen one-qubit rotation together with
the \emph{CNOT (Controlled NOT)} operation, which acts as
\begin{equation}
|x,y\rangle \mapsto |x,y\oplus x\rangle\;,
\end{equation}where $\oplus$ means addition modulo
2 \cite{PreskillNotes}.
Like in the classical case there are infinitely many sets of
universal gates. Notably,
also any generic (i.e., randomly
chosen) two-qubit gate (together with the possibility of
switching the leads in order to swap qubits)
is itself a universal set, very much like
the NAND gate is for classical computing \cite{Gates} \footnote{Any such generic
quantum gate has so-called entangling power \cite{Collins}, in
that it may transform a product state vector into one that can no
longer be written as a tensor product. Such quantum mechanical
pure states are called {\em entangled}. In intermediate steps of a
quantum algorithm the physical state of the system is in general
highly multi-particle entangled. In turn, the implementation of
quantum gates in distributed quantum computation requires
entanglement as a resource \cite{Dist}.}. Notably,
any quantum circuit that makes use of a certain universal set of
quantum gates can be simulated by a different quantum circuit
based on another universal set of gates with only polylogarithmic
overhead \cite{KitaevSim,Solovay,Mike}.
A particularly
useful single-qubit gate is the \emph{Hadamard gate},  acting as
\begin{eqnarray}
    |0\rangle&\mapsto&H |0\rangle=   (|0\rangle +
|1\rangle)/\sqrt{2},\,\,\,\,    |1\rangle\mapsto  H|1\rangle=
(|0\rangle - |1\rangle)/\sqrt{2}.
\end{eqnarray}
A \emph{phase gate} does nothing but multiplying one of the basis
vectors with a phase,
\begin{eqnarray}
    |0\rangle&\mapsto&  |0\rangle  ,\,\,\,\,    |1\rangle\mapsto
i  |1\rangle ,
\end{eqnarray}
and a \emph{Pauli gate} corresponds to one of the three unitary Pauli
matrices (see Fig.\ \ref{fig2}).
The CNOT, the Hadamard, the phase gate, and the Pauli gate
are quantum gates of utmost importance.
Given their key status in many quantum algorithms, one might be
tempted to think that
with these ingredients alone (together with measurements of Pauli operators,
see below),
powerful quantum algorithms may be constructed that outperform
the best known classical algorithm to a problem. This intuition is
yet not correct: it is the content
of the \emph{Gottesman-Knill theorem} that any quantum circuit
consisting of only these ingredients
can be simulated efficiently  on a classical computer
\cite{Mike,Gottesman}.
The proof of the Gottesman-Knill-theorem is deeply rooted in the
stabilizer formalism that we will
encounter later in the context of quantum error correction.

One of the crucial differences between classical and quantum
circuits is, that in the quantum case the COPY operation is not
possible. In fact, the linearity of quantum mechanics forbids a
device which copies an unknown quantum state -- this is known as
the \emph{no-cloning theorem}\footnote{If one knows that the state
vector is either $|0\rangle$ or $|1\rangle$ then a cloning machine
is perfectly consistent with rules of quantum mechanics. However,
producing perfect {\em clones} of an arbitrary quantum state as given by
Eq.\  (\ref{Psi}) is prohibited, as has been shown by
Wootters, Zurek, and Dieks \cite{WZDCloning}.}. The
latter has far-reaching consequences, of which the most prominent
one is the possibility of quantum cryptography coining this
``no-go theorem'' into an application \cite{Crypto}.

\subsection{How to program a quantum computer?} The good thing about the
classical computer on which this chapter has been
written is that it is
programmable. It is a single device capable of performing
different operations depending on the program it is given: word
processing, algebraic transformations, displaying movies, etc.. To
put it in more abstract words a classical computer is a
\emph{universal gate array}: we can program \emph{every} possible
function with $n$ input and $n$ output bits by specifying a
program of length $n2^n$. That is, a fixed circuit with
$n(1+2^n)$ input bits can be used in order to compute any function
on the first $n$ bits in the register. Is the same true for
quantum computers? Or will these devices typically be
made-to-measure with respect to a single task?

Nielsen and Chuang showed that quantum computers cannot be
universal gate arrays \cite{Programmable}.
Even if the program is itself given in form
of a quantum state it would require a program register of infinite
length in order to perform an arbitrary (unitary) operation on a
finite number of qubits -- universality was shown to be only
possible in a probabilistic manner. In this sense, quantum
computers will not be the kind of all purpose devices which
classical computers are. In practice, however, any finite set of
quantum programs can run on a quantum computer with a finite
program register. This issue applies, however, to the
programming of a  quantum computer with a fixed hardware,
which is, needless to say,
still in the remote future
as a physical device.

\subsection{Quantum error correction}

When it comes to experimental realizations
of quantum computers, we will have to deal with errors in the
operations, and we will have to find a way to protect the
computation against these errors: we have to find a way of doing
\emph{error correction}. Roughly speaking,
error correction in classical computers is essentially based on two
facts:
\begin{itemize}
\item[(i)] Computing with classical bits itself provides a simple way of
error correction in the form of a \emph{lock-in-place mechanism}.
If the two bits are for instance realized by two different
voltages (like it is the case in our computers as well as in our
brains) then the difference can simply be chosen large enough such
that typical fluctuations are small compared to the threshold
separating the two bits.

\item[(ii)] The information can be copied and then stored or
processed in a redundant way. If, for instance, an error occurs
to one of three
copies of a bit, we can recover the original information by
applying a majority vote. Of course, there are much more refined
versions of this method.

\end{itemize}
Unfortunately, in quantum computers we can not use either of these
ideas in a straightforward manner:
\begin{itemize}
\item[(i)] There is no lock-in-place mechanism and
\item[(ii)] the no-cloning
theorem forbids to copy the state of the qubits.
\end{itemize}
Naively measuring the state of the system to find out what error
has actually happened before correcting it
does not help, as any such attempt would necessarily
disturb the state in an irreversible manner.
So it was at the very
beginning of quantum information science not clear whether or not
under physically reasonable assumptions fault tolerant quantum
computing would be possible. It was obvious from the beginning on, in
turn, that there would be a need to achieve the goal of suitable
quantum error correction in some way. Without appropriate error correction
techniques, the promise of the Shor class
quantum computer as a computational device potentially
outperforming modern classical computers  could quite certainly not be met.

Fortunately, it could be shown by Steane, Shor,
and many other
researchers that error correction is nevertheless possible
and that the above problems can indeed be overcome
\cite{Steane,Steane2,ShorError,Laflamme}.
The basic idea is  that a logical
qubit can be protected by encoding it in a non-local manner into
several physical qubits.
This amounts to a lossless encoding
in longer code words to make the states robust against the effects of noise,
without the need of actually copying the quantum state under consideration and
introducing redundancy in the literal sense.

\section{Elementary quantum algorithms}\label{SecDeutsch}

In the same scientific paper in which David Deutsch introduced
the notion of the
universal quantum computer,
he also presented the first quantum algorithm
\cite{Deutsch85}\footnote{Quantum Turing machines were
first considered by Benioff \cite{Benioff} and developed
by Deutsch in Ref.\ \cite{Deutsch85}.}.
The problem that this
algorithm addresses,
later referred to as Deutsch's problem, is a very simple one. Yet the
\emph{Deutsch algorithm} already exemplifies
the advantages of a quantum computer through skillfully
exploiting quantum parallelism.
Like the Deutsch algorithm, all other elementary quantum algorithms
in this section
amount to deciding which {\em black box}
out of finitely many alternatives
one has at hand. Such a black box is often also referred
to as {\em oracle}.
 An input may be given to the oracle,  one may read out or use the outcome
 in later steps of the quantum algorithm, and
the objective is to find out the functioning of the black box.
It is assumed that this oracle operation can be implemented
with some sequence of quantum logic gates.
The complexity of
the quantum algorithm is then quantified in terms of the number of
queries to the oracle.

\subsection{Deutsch algorithm} With the help
of this algorithm, it is possible to decide whether a function
has a certain property
with a single call of the function,
instead of two calls that are necessary classically.
Let
\begin{equation}
    f:\{0,1\}\longrightarrow\{0,1\}
\end{equation}
be a function that has  both a one-bit domain and range. This
function can be either {\em constant} or {\em balanced}, which
means that either $f(0)\oplus f(1)=0$ or $f(0)\oplus f(1)=1$
holds.
%
%
The problem is to find
out with the minimal number of
function calls whether this function $f$ is constant or balanced. In
colloquial terms,
the problem under consideration may be described as a procedure to
test whether a coin is
fake (has two heads or two tails) or a genuine coin.

\begin{figure}
\centerline{
\includegraphics[width=4.2cm]{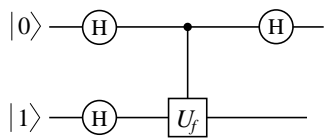}
}
    \caption{The circuit of the Deutsch algorithm.}\label{fig3}
\end{figure}

Classically, it is obvious that two function calls are required to
decide which
of the two allowed cases is realised, or, equivalently, what the value
of $f(0)\oplus f(1)$ is. A way to compute the function $f$ on a
quantum computer
is to transform the state vector of two qubits
according to
\begin{equation}\label{oracle}
    |x,y\rangle \mapsto U_f |x,y\rangle =  |x,f(x)\oplus y\rangle.
\end{equation}
In this manner, the evaluation can be realized
unitarily. 
The above map
is what is called a standard quantum oracle (as
opposed to a minimal quantum oracle \cite{Minimal},
which would be of the form
$|x \rangle \mapsto |f(x) \rangle$).
The claim now is that using
such an oracle, a single function call is sufficient for the
evaluation of $f(0)\oplus f(1)$. In order to show this, let us assume that we have
prepared
two qubits in the state with state vector
\begin{equation}
    |\Psi\rangle = (H\otimes H)|0,1\rangle,
\end{equation}
where $H$ denotes the Hadamard gate of section \ref{QvsC}.
We now apply the unitary $U_f$ once
to this state, and finally apply another Hadamard gate to the first
qubit. The resulting state
vector hence reads as (see Fig.\ \ref{fig3})
\begin{equation}
    |\Psi'\rangle = (H\otimes \id) U_f (H\otimes H)|0,1\rangle.
\end{equation}
A short calculation shows that $|\Psi'\rangle$ can be evaluated to
\begin{equation}
    |\Psi'\rangle = \pm |f(0)\oplus f(1)\rangle  H |1\rangle.
\end{equation}
The second qubit is in the state corresponding to the vector $H
|1\rangle$, which is
of no relevance to our problem. The state of the first qubit,
however, is quite remarkable:
encoded is $|f(0)\oplus f(1)\rangle$, and both alternatives are
decidable with unit probability
in a measurement in the computational basis,
as the two state vectors are
orthogonal\footnote{Note that the presented
algorithm is not quite the one in the original paper by Deutsch,
which allowed for an inconclusive
outcome in the measurement. This deterministic version of the Deutsch
algorithm is due to
Cleve, Ekert, Macchiavello, and Mosca \cite{Mosca}.}.
That is, with a single
measurement of the state, and notably, with a single call of the
function $f$, of the
first qubit we can decide whether $f$ was constant or balanced.

\subsection{Deutsch-Jozsa algorithm}
The Deutsch algorithm
has not yet any implication on superiority of a quantum computer
as compared
to a classical as far as the query complexity
is concerned. After all, it is
merely one function call instead of
two. The situation is different in case of the extension of the
Deutsch algorithm known as
\emph{Deutsch-Jozsa algorithm} \cite{DeutschJozsa}.
Here, the task is again
to find out whether a function is constant
or balanced, but  $f$ is now a function
\begin{equation}
    f: \{0,1\}^N \longrightarrow \{0,1\},
\end{equation}
where $N$ is some natural number. It is promised that the function is
either constant, which now
means that either $f(i)=0$ for all
$i=0,...,2^N-1$ or $f(i)=1$ for all $i$, or balanced.
It is said to be balanced if
the image under $f$ takes as many times the value $1$ as the value
$0$.
The property to be balanced or constant can be said to be a global
property of
several function values. It
is a promised
property of the function, which is why the Deutsch-Jozsa algorithm is being
classified as
a \emph{promise algorithm}. There are only two possible black boxes
to the disposal, and the tasks is to find out which one is realised.

\begin{figure}
\centerline{
\includegraphics[width=4.5cm]{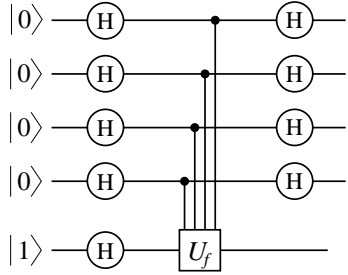}
}
    \caption{The circuit of the Deutsch-Jozsa algorithm.}\label{fig4}
\end{figure}

It is clear how many
times one needs to call the function on a classical computer:
the worst case scenario is that after $2^N/2$ function calls,
the answer has been always $0$ or always $1$.  Hence, $2^N/2+1$
function calls are required to know
with certainty whether the function is balanced or constant (a result
that can  be significantly
improved if probabilistic algorithms are allowed for).
Quantum mechanically, again a single
function call is sufficient. Similarly to the above situation, one
may prepare $N+1$ qubits in the state
with state vector
\begin{equation}
    |\Psi\rangle = H^{\otimes (N+1)}
    |0,...,0,1\rangle,
\end{equation}
and apply the unitary $U_f$ as in Eq.\ (\ref{oracle})
to it, acting as an oracle, and apply
$H^{\otimes N}\otimes \id$ to the resulting state, to obtain
(see Fig.\ \ref{fig4})
\begin{equation}
    |\Psi'\rangle
    =(H^{\otimes N}\otimes \id) U_f H^{\otimes (N+1)}
    |0,...,0,1\rangle.
\end{equation}
In the last step, one performs a  measurement on the first
$N$ qubits in the computational basis.
In effect, one observes that exactly if the function $f$ is
constant, one obtains the measurement outcome corresponding to
$|0,0,...,0\rangle$ with certainty. For any other output the
function was balanced. So again, the test for the promised
property can be performed with a single query, instead of
$2^N/2+1$ classically \footnote{ There are a number of related
problems that show very similar features. In the
\emph{Bernstein-Vazirani algorithm} \cite{Bernstein} again a
function
 $f:\{0,1\}^N\longrightarrow \{0,1\}$ is given, promised to be of the
form
\begin{equation}
f(x)=a x.
\end{equation}
for $a,x\in \{ 0,1\}^N$ for some natural number $N$. $a x$ denotes
the standard scalar product $a x=a_0 x_0+...+ a_{2^N-1}
x_{2^N-1}$. How many measurements are required to find the vector
$a$ of zeros and ones? Classically, one has to perform
measurements for all possible arguments, and in the end solve a
system of linear equations. With the standard oracle
$|x,y\rangle\mapsto |x,f(x)\oplus y\rangle$ at hand, in its
quantum version in the Bernstein-Vazirani algorithm only a single
call of the oracle is required. Although it has been convincingly
argued that one does not have to evoke the metaphor of quantum
parallelism to interpret the functioning of the quantum computer
in the Bernstein-Vazirani problem -- the difference from quantum
to classical lies rather in the ability to reverse the action of a
CNOT gate by means of local operations on the control and target
qubits -- the surprisingly superior performance of the quantum
algorithm to its classical counterpart is manifest.}.

After all, the performance of the Deutsch-Jozsa
algorithm is quite impressive.
If there is a drawback to this, yet,
it is that unfortunately, the Deutsch-Jozsa algorithm is to some
extent
artificial in nature, and it lacks an actual practical application
emerging in a natural context. The astonishing difference in the number
of queries in the quantum and classical case also disappears if
classically probabilistic algorithms are allowed for: in fact, using a
probabilistic algorithm, a polynomial number of queries
achieves an exponentially good success
probability.

\subsection{Simon's algorithm}

{\em Simon's problem} is an instance of an oracle problem which is
classically hard,
 even for probabilistic algorithms, but tractable for
quantum computers \cite{Simon}.
The task is to find the period $p$ of a certain
function $f:\{0,1\}^N\longrightarrow\{0,1\}^N$, which is promised to
be 2-to-1 with $f(x)=f(y)$ if and only if $y=x\oplus p$.
Here, $x$ and $y$ denote binary words of
length $N$, where
$\oplus$ now means bitwise addition modulo 2. The problem can be stated as
a decision problem as well and the goal would then be to decide
whether or not there is a period, i.e., whether $f$ is 2-to-1 or
1-to-1.

Classically the problem is hard, since the probability of having
found two identical elements $x$ and $y$ after $2^{N/4}$ queries
is still less than $2^{- N/2}$. Simon's quantum solution is now
the following: start with a state vector $(H |0\rangle)^{\otimes
N}|0\rangle^{\otimes N}$ and run the oracle once yielding the
state vector $2^{- N/2}\sum_x |x\rangle |f(x)\rangle$. Then
measure the second register\footnote{Note that this step is not
even necessary -- it is merely a pedagogical one.}. If the
measurement outcome is $f(x_0)$, then the state vector of the
first register will be
\begin{equation} \label{eqSim1st} \frac1{\sqrt{2}}\bigl(|x_0\rangle
+ |x_0 \oplus  p\rangle \bigr).
\end{equation} Applying a Hadamard gate to each of the $N$ remaining qubits
leads to
\begin{eqnarray}
&&\frac1{{2^{(N+1)/2}}} \sum_y \left( (-1)^{x_0\cdot y}+
(-1)^{(x_0\oplus p)\cdot y} \right) |y\rangle\\
&&=\frac1{{2^{(N-1)/2}}} \sum_{p\cdot y=0} (-1)^{x_0\cdot y}
|y\rangle.
\end{eqnarray} If we finally measure the first register in
computational basis, we obtain a value $y$ which is such that
$y\cdot p=0$ modulo 2. Repeating this procedure in order to get
$N-1$ linearly independent vectors $y_1,\ldots, y_{N-1}$ we can
determine $p$ from the set of equations $\{y_i\cdot p=0\}$. To
this end we have to query the oracle ${\cal O}(N)$
times\footnote{This symbol is the ``big-O'' Landau symbol for the
asymptotic upper bound. In the rest of the chapter, this notation
will be used even if the asymptotic behavior could be specified
more precisely.}. Hence, we get an exponential speed up compared
to any classical algorithm. And in contrast to the Deutsch-Jozsa
algorithm this exponential gap remains if we allow for
probabilistic classical algorithms\footnote{Simon's problem is an
instance of an oracle problem relative to which BPP$\neq$BQP. That
is, classical and quantum polynomial-time complexity classes for
bounded error probabilistic algorithms differ relative to Simon's
problem.}.
Simon's algorithm has much in common with Shor's algorithm: they
both try to find the period of a function\footnote{Whereas Simon's
problem is to find a period in $(\mathbb{Z}_2)^N$, Shor's
algorithm searches for one in $\mathbb{Z}_{2^N}$.}, both yield an
exponential speed-up, and both make use of classical algorithms in
a post processing step. Actually, Shor's work was inspired by
Simon's result.

\section{Grover's database search algorithm}

The speed-up of the presented quantum algorithms for the
Deutsch-Jozsa and Simon's problem is enormous. However, the oracle
functions are constrained to comply with certain promises and the
considered tasks hardly appear in practical applications. In
contrast to that, Grover's algorithm deals with a frequently
appearing problem \cite{Grover}: \emph{database search.}

Assume we have an unsorted list and want to know the largest
element, the mean, whether there is an element with certain
properties or the number of such elements -- all these are common
problems or necessary subroutines for more complex programs and
due to Grover's algorithm all these problems in principle admit a
typically quadratic speed-up compared to classical solutions. Such
an improvement of the performance might not appear very
spectacular, however, the problems to which it is applicable are
quite numerous\footnote{For instance, the standard solution to all
NP-complete problems is doing an exhaustive search. Hence,
Grover's algorithm would speed-up finding a solution to the
travelling salesman, the Hamiltonian cycle and certain coloring
problems.} and the progress from the ordinary Fourier transform to
the FFT already demonstrated how a quadratic speed-up in an
elementary routine can boost many applications.

Consider the problem of searching a marked element
$x_0\in\{1,\ldots,N\}$ within an unsorted database of length $N
=2^n$. Whereas classically we have to query our database ${\cal
O}(N)$ times in order to identify the sought element, Grover's
algorithm will require only ${\cal O}(\sqrt{N})$ trials. Let the
database be represented by a unitary\footnote{$|x_0\rangle\langle
x_0|$ means the projector onto the vector $|x_0\rangle$. That is
$U_{x_0}|x\rangle = (-1)^{\delta_{x,x_0}}|x\rangle$.}
\begin{equation}\label{eqGrovOracle} U_{x_0}= \id - 2 |x_0\rangle\langle
x_0|,
\end{equation} which  flips the sign of  $|x_0\rangle$ but
preserves all vectors orthogonal to $|x_0\rangle$. The first step
of the algorithm is to prepare an equally weighted superposition
of all basis states $|\Psi\rangle =\frac1{\sqrt{N}}\sum_x
|x\rangle$. As we have already seen previously this can be
achieved by applying $N$ Hadamard gates to the state vector
$|0\rangle$. Next, we apply the \emph{Grover operator} \begin{equation}
\label{eqGrovOp}G= U_{\Psi}U_{x_0}\ ,\quad U_{\Psi}=
2|\Psi\rangle\langle\Psi|-\id
\end{equation}to the state vector $|\Psi\rangle$. Geometrically, the
action of $G$ is to rotate $|\Psi\rangle$ towards $|x_0\rangle$ by
an angle $2\varphi$ where $\sin\varphi=
|\langle\Psi|x_0\rangle|=1/\sqrt{N}$. The idea is now to iterate
this rotation $k$-times until the initial state is close to
$|x_0\rangle$, i.e., \begin{equation} G^k|\Psi\rangle \approx
|x_0\rangle.
\end{equation}Measuring the system (in computational basis) will then reveal the value of
$x_0$ with high probability.

So, how many iterations do we need?
Each step is a $2\varphi$-rotation
 and the initial  angle between
$|\Psi\rangle$ and $|x_0\rangle$ is
$\pi/2-\varphi$.\footnote{This clarifies why we start with
the state vector
$|\Psi\rangle$: the overlap $|\langle\Psi|x_0\rangle|$
does not depend on $x_0$.} Using that for large $N$
$\sin\varphi\approx\varphi$ we see that $k\approx
\pi \sqrt{N}/4$ rotations will do the job and the probability
of obtaining a measurement outcome different from $x_0$ will
decrease as ${\cal O}(1/N)$. Since every step in the Grover
iteration queries the database once, we need indeed only ${\cal
O}(\sqrt{N})$ trials compared to ${\cal O}(N)$ in classical
algorithms. To exploit this speed-up we need of course an
efficient implementation not only of the database-oracle $U_{x_0}$,
but also of the unitary $U_{\Psi}$. Fortunately, the latter can
be constructed out of ${\cal O}(\log N)$ elementary gates.

What if there are more than one, say $M$, marked elements? Using
the equally weighted superposition of all the respective states
instead of $|x_0\rangle$ we can essentially repeat the above
argumentation and obtain that ${\cal O}(\sqrt{N/M})$ queries are
required in order to find one out of the $M$ elements with high
probability. However, performing further Grover iterations would
be overshooting the mark: we would rotate the initial state beyond
the sought target and the probability for finding a marked element
would rapidly decrease again. If we initially do not know the
number $M$ of marked elements this is, however, not a serious
problem. As long as $M\ll N$ we can still gain a quadratic
speed-up by simply choosing the number of iterations randomly
between 0 and $\pi \sqrt{N}/4$. The probability of finding a
marked element will then be close to $1/2$ for every $M$.
Notably, Grover's algorithm is optimal in the sense that any quantum
algorithm for this problem will necessarily require
${\cal O}(\sqrt{N/M})$ queries \cite{GroverOptimal}.
%

\section{Exponential speed-up in Shor's factoring
algorithm}\label{SecShor}

Shor's  algorithm \cite{Shor}
is without doubt not only one of the
cornerstones of quantum information theory but also one of the
most surprising advances in the theory of computation itself: a
problem, which is widely believed to be {\emph{hard}} becomes
\emph{tractable} by refereing to (quantum) physics -- an approach
completely atypical for the theory of computation, which usually
abstracts away from any physical realization.

The problem Shor's algorithm deals with is \emph{factorization}, a
typical NP problem.
Consider for instance the task of finding
the prime factors of 421301. With pencil and paper it might
probably take more than an hour to find them. The inverse problem,
the multiplication $601\times 701$, can, however, be solved in a
few seconds even without having pencil and paper at
hand\footnote{Actually, it takes eleven seconds for a randomly
chosen Munich schoolboy at the age of twelve (the sample size was
one).}. The crucial difference between the two tasks
multiplication and factoring is, however, how the degree of
difficulty increases with the length of the numbers. Whereas
multiplication belongs to the class of ``tractable'' problems for
which the required number of elementary computing steps increases
polynomially with the size of the input, every known classical
factoring algorithm requires an exponentially increasing number of
steps. This is what is meant by saying that factoring is an
``intractable'' or ``hard'' problem. In fact, it is this
discrepancy between the complexity of the factoring problem and
its inverse which is exploited in the most popular public key
encryption scheme based on RSA -- its security heavily relies on the
assumed difficulty of factoring.
In a nutshell the idea of Shor's factoring algorithm is the
following:
\begin{itemize}
    \item[(i)]
    \emph{Classical part:} Using some elementary number theory one can show
that the problem of finding a factor of a given integer is
essentially equivalent to determining the period of a certain
function.

    \item[(ii)] \emph{QFT for period-finding}: Implement the function
    from step (i) in a quantum circuit and apply it to a superposition of all
    classical input
    states. Then perform  a discrete   quantum
    Fourier transform (QFT) and measure the output. The measurement outcomes
    will be probabilistically distributed according to the
    inverse of the sought period. The latter can thus be determined (with certain probability) by
    repeating the procedure.

    \item[(iii)] \emph{Efficient implementation:} The crucial point of
    the algorithm is that the QFT as well as the function from
    step (i) can be efficiently implemented, i.e., the number of required
    elementary operations grows only polynomially with the size of the input.
    Moreover, the probability of success of the algorithm can be made arbitrary
    close to one without exponentially increasing effort.

\end{itemize}

Clearly, the heart of the algorithm is an efficient implementation
of the QFT. Since  Fourier transforms enter in many mathematical
and physical problems one might naively expect an exponential
speedup for all these problems as well. However, the outcome of
the QFT is not explicitly available but ``hidden'' in the
amplitudes of the output state, which can not be measured
efficiently. Only global properties of the function, like its
period, can in some cases
be determined efficiently.

Nevertheless, a couple of other applications are known for which
the QFT leads again to an exponential speed up compared to the
known classical algorithms. The abstract problem, which
encompasses all these applications is known as the ``hidden
subgroup problem'' \cite{Mike}
and another rather prominent representative of
this type is the discrete logarithm problem.
Let us now have a more detailed look at the ingredients for Shor's
algorithm.

\subsection{Classical part}

Let $N$  be an odd  number we would like to factor and $a<N$ an
integer which has no non-trivial  factor in common with $N$, i.e.,
$gcd(N,a)=1$. The latter can efficiently be checked by Euclid's
algorithm\footnote{in ${\cal O}((\log N)^3)$ time.}. A factor of
$N$ can then be found indirectly by determining the period $p$ of
the function $f:{\mathbbm Z} \longrightarrow {\mathbbm Z}_N$ defined as
\begin{equation}\label{funcfp} f(x)= a^x
\;\mbox{mod}{N}.
\end{equation}Hence, we are looking for a solution of the
equation $ a^p-1  =
0 \;\mbox{mod}{N}. $ Assuming $p$ to be
even we can decompose
\begin{equation} a^p-1=(a^{\frac{p}2}+1)(a^{\frac{p}2}-1)=
 0 \;\mbox{mod}{N},
\end{equation}and therefore either one or both terms $(a^{\frac{p}2}\pm
1)$ must have a factor in common with $N$. Any non-trivial common
divisor of $N$ with $(a^{\frac{p}2}\pm 1)$, again calculated by
Euclid's algorithm, yields thus a non-trivial factor of $N$.

Obviously, the described procedure is only successful if $p$ is
even and the final factor is a non-trivial one. Fortunately, if we
choose $a$ at random\footnote{For each randomly chosen $a$ we have
again to check whether $gcd(N,a)=1$. The probability for this can
be shown to be larger than $1/\log N$. The total probability of
success is thus at least $1/(2 \log N)$.}, this case occurs with
probability larger than one half unless $N$ is a power of a prime.
The latter can, however, be checked again efficiently by a known
classical algorithm, which returns the value of the prime.
Altogether a polynomial time algorithm for determining the period
of the function in Eq.\ \ (\ref{funcfp}) leads to a probabilistic
polynomial time algorithm which either returns a factor of $N$ or
tells us that $N$ is prime.

\subsection{Quantum Fourier Transform}

The step from the ordinary discrete Fourier transform (based on
matrix multiplication) to the Fast Fourier Transform (FFT) has
been of significant importance for signal and image processing as
well as for many other applications in scientific and engineering
computing\footnote{Although FFT is often attributed
to
Cooley
and
Tukey in 1965, it is now known that around 1805
Gauss used the algorithm already to interpolate the trajectories
of asteroids \cite{Gauss}.}. Whereas the naive way of calculating
the discrete Fourier transform
\begin{equation}
\hat{c}_y=\frac1{\sqrt{n}}\sum_{x=0}^{n-1} c_x e^{\frac{2\pi i}n
xy}
\end{equation}by matrix multiplication takes ${\cal O}(n^2)$ steps,
the FFT requires ${\cal O}(n\log n)$.
The {\em quantum Fourier transform} (QFT) 
\cite{Shor,Coppersmith,Kitaev95,Beth}
is in fact a straightforward
quantum generalization of the FFT, which can, however, be
implemented using only ${\cal O}((\log n)^2)$ elementary
operations -- an exponential speedup!

\begin{figure}
\centerline{
\includegraphics[width=6.5cm]{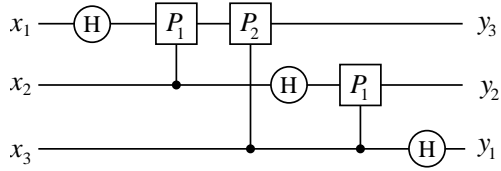}
}
    \caption{The circuit of a discrete quantum Fourier
    transform on three qubits. The gate $P_d$ adds a
    conditional relative phase $\pi/2^d$, where $d$ is the
    distance between the two involved qubits in the circuit.}\label{figQFT}
\end{figure}

Let now the computational basis states of $q$ qubits
be characterized by the binary representation of numbers
$x=\sum_{i=1}^q x_i 2^{i-1}$ via
\begin{equation}
|x\rangle = |x_1,\ldots, x_q\rangle.
\end{equation}
That is, in this subsection $x$ denotes from now on
a natural number
or zero and not a binary word.
Then for $n=2^q$ the QFT acts on a general state vector of
$q$ qubits as $ \sum_x c_x |x\rangle \mapsto \sum_y \hat{c}_y
|y\rangle . $ This transformation can be implemented using only
two types of gates: the Hadamard gate and conditional phase gates
$P_d$ acting as
\begin{equation}
|a,b\rangle \mapsto |a,b\rangle e^{\delta_{a+b,2}\pi i /2^d}.
\end{equation}
which rotate
the relative phase conditionally by an angle $\pi 2^{-d}$, where
$d$ is the ``distance'' between the two involved
qubits.


Fig. \ref{figQFT} shows the quantum circuit, which implements the
QFT on $q=3$ qubits. The extension of the circuit to more than
three qubits is rather obvious and since $q(q+1)/2$ gates are
required its complexity is ${\cal O}(q^2)={\cal O}((\log n)^2)$.
Being only interested in an approximate QFT we could reduce the
number of gates even further to ${\cal O}(\log n)$ by dropping all
phase gates $P_d$ with $d\geq m$. Naturally, the accuracy  will
then depend on $m$.\footnote{An $\epsilon$-approximation of the
QFT (in the 2-norm) would require ${\cal O}(q \log(q/\epsilon))$
operations, i.e., $m$ is of the order $\log(q/\epsilon)$ (cf.
\cite{Coppersmith}).}

\subsection{Joining the pieces}

Let us now sketch how the QFT can be used to compute the period
$p$ of the function in Eq.\  (\ref{funcfp}) efficiently. Consider two
registers of $q$ qubits each, where $2^q=n\geq N^2$ and all the
qubits are in the state vector $|0\rangle$ initially. Applying a Hadamard
gate to each qubit in the first register yields
$\sum_x |x,0\rangle/{\sqrt{n}}$. Now suppose we have
implemented the function in Eq.\  (\ref{funcfp}) in a quantum circuit
which acts as $|x,0\rangle\mapsto |x,f(x)\rangle$, where $x$
is taken from ${\mathbbm Z}_n$.
Applying this
to the state vector
and then performing a QFT on the first register we
obtain
\begin{equation}\label{Shorfinalstate}
\frac1n\sum_{x,y=0}^{n-1} e^{\frac{2\pi i}n xy}
|y,f(x)\rangle.\end{equation} How will the distribution of
measurement outcomes look like if we now measure the first
register in computational basis? Roughly speaking, the sum over
$x$ will lead to constructive interference whenever $y/n$
is close to a multiple of the inverse of the period $p$ of $f$ and
it yields destructive interference otherwise. Hence, the
probability distribution for measuring $y$ is sharply peaked
around multiples of $ n /p$ and $p$ itself can be determined
by repeating the whole procedure  ${\cal O}(\log N)$ times\footnote{For the cost of more classical 
postprocessing it is even possible to reduce the expected number of 
required trials to a constant (cf.\ \cite{Shor}).}. At the
same time the probability of success can be made arbitrary close
to one. In the end we can anyhow easily verify whether the result,
the obtained factor of $N$, is valid or not.

What remains to be shown is that the map
\begin{equation}
|x,0\rangle\mapsto |x,f(x)\rangle\ ,\quad f(x)=a^x \;\mbox{mod}{N}
\end{equation} can be implemented efficiently. This can be done
by repeatedly squaring in order to get $a^{2^j}$ mod $N$ and then
multiplying a subset of these numbers according to the binary
expansion of $x$. This requires ${\cal O}(\log N)$ squarings and
multiplications of $\log N$-bit numbers. For each multiplication
the ``elementary-school algorithm'' requires ${\cal O}((\log
N)^2)$ steps. Hence, implementing this simple classical algorithm
on our quantum computer we can compute $f(x)$ with ${\cal O}((\log
N)^3)$ elementary operations. In fact, this part of performing a
standard classical multiplication algorithm on a quantum computer
is the bottleneck in the quantum part of Shor's algorithm. If
there would be a more refined \emph{quantum modular
exponentiation} algorithm we could improve the asymptotic
performance of the algorithm\footnote{In fact, modular
exponentiation can be done in ${\cal O}((\log N)^2 \log\log N
\log\log\log N)$ time by utilizing the Sch\"onhagen-Strassen
algorithm for multiplication \cite{Strassen}.
However, this is again a classical
algorithm, first made reversible and then run on a quantum
computer. If there exists a faster quantum algorithm it would even
be possible that breaking RSA codes on a quantum computer is
asymptotically faster than the encryption on a classical
computer.}.

Altogether, the quantum part of Shor's factoring algorithm
requires of the order $(\log N)^3$ elementary steps, i.e., the
size of the circuit is cubic in the length of the input. As
described above, additional classical preprocessing and
postprocessing is necessary in order to obtain a factor of $N$.
The time required for the classical part of the algorithm is,
however, polynomial in $\log N$ as well, such that the entire
algorithm does the job in polynomial time. In contrast to that,
the running time of the number field sieve, which is currently the
best classical factoring algorithm, is $\exp[{\cal
O}((\log N)^{\frac13}(\log\log N)^{\frac23})]$.
Moreover, it is widely believed that factoring is a classically
hard problem, in the sense that there exists no classical
polynomial time algorithm. However, it is also believed that
proving the latter conjecture (if it is true) is extremely hard
since it would solve the notorious $P\stackrel{?}{=}NP$ problem.

\section{Adiabatic quantum computing}\label{SecAQC}

Shor's factoring algorithm
falls into a certain class of quantum algorithms, together with
many other important algorithms, such as the algorithm for
computing orders of solvable groups \cite{Watrous} and
the
%
%
efficient quantum algorithm for finding solutions of
Pell's equation \cite{Hallgren}:
it is an instance of a   hidden subgroup problem.
In fact, it has turned out in recent years
that it appears difficult to
leave the framework of hidden subgroup problems, and
to find novel quantum algorithms  for
practically relevant problems. This motivates the quest
for entirely new approaches to finding such new algorithms.
The algorithm of Ref.\ \cite{Childs}
based on {\em quantum random walks} \cite{Walks}
is an important example of such a new approach, although the
problem it solves does not appear in a particularly
practical context. Another approach is the framework of
adiabatic quantum algorithms:

In 2000, Farhi, Goldstone, Gutmann, and Sipser introduced a new
concept to the study of quantum algorithms, based on the adiabatic theorem
of quantum mechanics \cite{Farhi1}.
The idea is the following:  let
$f:\{0,1\}^N\longrightarrow {\mathbbm R}$ be a cost function of which we would like
to find the global minimum, assumed in $x\in\{0,1\}^N$.
In fact, any local combinatorical search problem can be formulated
in this form. For simplicity, suppose that this global minimum is
unique. Introducing the {\em problem Hamiltonian}
\begin{equation}
    H_T =\sum_{z\in\{0,1\}^N}
    f(z) |z\rangle\langle z|,
\end{equation}
the problem of finding the $x\in\{0,1\}^N$
where $f$ attains its minimum
amounts to identifying the eigenstate $| x\rangle$
of $H_T$ corresponding
to the smallest eigenvalue $f(x)$, i.e., the {\em ground state energy}
associated with $H_T$.
But how does one find the ground state in the first place?
The key idea is to consider another Hamiltonian $H_0$, with the property
that the system can easily be prepared in its ground state,
which is again assumed to be unique. One then
interpolates between the two Hamiltonians, for example
linearly
\begin{equation}
    H(t)= \frac{t}{T} H_T + \bigl(1- \frac{t}{T}\bigr)
    H_0,
\end{equation}
with $t\in[0,T]$, where $T$ is the {\em run time}
of the adiabatic quantum algorithm. This Hamiltonian governs
the time evolution of the quantum state of the system from
time $t=0$  until $t=T$.
According to the Schr{\"o}dinger equation,  the state vector evolves
as  $i \partial_t |\Psi(t)\rangle = H(t) |\Psi(t)\rangle $.
In a last step one performs a measurement
in the computational basis. If one obtains the
outcome associated with $|x \rangle$, then the measurement result
is just $x$, the minimal value of the function $f$. In this case
the probabilistic
algorithm is successful, which happens with
the {\em success probability}  $p=|\langle  x | \Psi(T)\rangle|^2$.
%
%

What are the requirements for such an algorithm to work, i.e., to
result in $x$ with a large success probability? The answer to this
question is provided by the {\em quantum adiabatic theorem}:
If the Hamiltonian $H(t)$ exhibits a non-zero spectral gap between
the smallest and the second-to-smallest eigenvalue for all $t\in[0,T]$,
then
the final state vector $|\Psi(T)\rangle$ will be close
to the state vector $|x\rangle$ corresponding to the ground state of $H_T$,
if the interpolation happens sufficiently
slowly, meaning that $T$ is sufficiently large.
The initial state is then said to be adiabatically transferred
to arbitrary accuracy into the desired ground state of the problem
Hamiltonian, which encodes the solution to the
problem. The typical problem of encountering local minima
that are distinct from the global minimum can in principle not even occur.
This kind of quantum algorithm is referred to as an
{\em adiabatic algorithm}.

Needless to say, the question is how large the time $T$ has to be chosen.
Let us denote with
\begin{equation}
    \Delta= \min_{t\in[0,T]} (E_t^{(0)}-  E_t^{(1)})
\end{equation}
the minimal spectral gap over the time interval $[0,T]$
between  the smallest $E_t^{(0)}$ and  the second-to-smallest eigenvalue $E_t^{(1)}$
of $H(t)$, associated with eigenvectors $|\Psi_t^{(0)}\rangle$ and
$|\Psi_t^{(1)}\rangle$, respectively,
and with
\begin{equation}
     \Theta=  T \max_{t\in[0,T]}
    |
    \langle
    \Psi_t^{(1)} | \partial_t H(t) | \Psi_t^{(0)}
    \rangle
     |
     = 
     \max_{t\in[0,T]}
    |
    \langle
    \Psi_t^{(1)} |H_T - H_0 | \Psi_t^{(0)}
    \rangle
     .
\end{equation}
Then, according to the quantum adiabatic theorem, the success
probability satisfies
\begin{equation}
    p= |\langle \Psi_{T}^{(0)}| \Psi(T)\rangle|^2 \geq 1-\varepsilon^2
\end{equation}
if
\begin{equation}
    T \varepsilon\geq \frac{\Theta}{\Delta^2}.
\end{equation}

The quantity $\Theta$ is typically polynomially bounded in $N$  for the
problems one is interested in, so the crucial issue is the behaviour of
the minimal gap $\Delta$. Time complexity is now quantified in terms of
the run time $T$ of the adiabatic algorithm. If one knew the spectrum of
$H(t)$ at all times, then one could immediately see how fast the algorithm
can be performed. Roughly speaking, the larger the gap, the faster the algorithm
can be implemented. The problem is that the spectrum of $H(t)$, which
can be represented as a $2^N\times 2^N$ matrix, is in general unknown.
Even to find lower bounds for the minimal spectral gap is extraordinarily
difficult, unless a certain
symmetry highly simplifies the problem of finding the spectrum.
After all, in order for the Hamiltonian to be
`reasonable', it is required that it is {\em local}, i.e., it is a sum of operators
that act only on a bounded number of qubits in $N$. This is a very
natural restriction, as it means that the physical interactions involve always
only a finite number of quantum systems \cite{Dorit}.
Note that an indication whether
the chosen run time $T$ for an adiabatic algorithm was appropriate, one
may start with the initial Hamiltonian and prepare the system in its
ground state, interpolate to the problem Hamiltonian and -- using the same interpolation --
back to the original Hamiltonian \cite{Cirac}.
A
necessary condition for the algorithm to have been successful is that finally,
the system is to a good approximation in the ground state of the initial Hamiltonian.
This is a method that should be accessible to an experimental implementation.

Adiabatic algorithms are known to reproduce the quadratic
speedup in the Grover algorithm for unstructured search problems
\cite{Cerf1}.
But adiabatic algorithms can also be applied to other instances of
search problems: In Ref.\ \cite{Annealing}
adiabatic algorithms have been compared with
simulated annealing algorithms, finding settings in which the
quantum adiabatic algorithm succeeded in polynomial time, but
for simulated annealing exponential time was necessary.
There is after all some numerical evidence that for structured
NP hard problems like MAX CLIQUE and 3-SAT, it may well be
that adiabatic algorithms offer an exponential speedup over the best
classical algorithm, again, assuming that $P\neq NP$
\cite{Farhi1}.
In fact, it can be shown that adiabatic algorithms can be
efficiently simulated on a quantum computer based on the
quantum circuit model, provided that the Hamiltonian is
local in the above sense (see also the subsequent section).
Hence, whenever an efficient
adiabatic algorithm can be found for a specific problem, this implies
an efficient quantum algorithm \cite{Dorit}.
The concept of adiabatic
algorithms may be a key tool to establish new algorithms beyond
the hidden subgroup problem framework.

\section{Simulating quantum systems}\label{SecQS}

A typical application of computers is that of being working horses
for physicists and engineers who want to simulate physical
processes and compute practically relevant properties of certain
objects from the elementary rules of physics. If many particles
are involved, the simulation might become cumbersome or even
impossible without exploiting serious approximations. This is true
classically as well as quantum mechanically\footnote{Even the type
of differential equation we have to solve can be very similar. The
classical diffusion equation is for instance essentially a real
version of the quantum mechanical Schr\"odinger equation.}:
simulating turbulences is not necessarily easier than dealing with
high temperature superconductors. There is, however, a crucial
difference between classical and quantum systems regarding how
many are ``many particles''. Whereas the dimension of the
classical phase space grows linearly with the number of particles,
the size of the quantum mechanical Hilbert space
increases exponentially. This implies that the exact simulation of an
arbitrary quantum system of more than 25 qubits is already no
longer feasible on today's computers. Consider for instance a
closed system of $N$ (say 25) qubits whose time evolution is
determined by a Hamiltonian $H$ via Schr{\"o}dinger dynamics,
\begin{equation}\label{eqHamiltonian}
|\Psi(t)\rangle = e^{-i H t}|\Psi(0)\rangle.
\end{equation}Since $H$ is a Hermitian $2^N\times 2^N$ matrix it
is, although often sparse, extremely hard to exponentiate -- for
$N=25$ it has about $10^{15}$ entries!

Once we have the building blocks for a \emph{universal} quantum
computer of $N$ qubits, i.e., a universal set of gates, we can in
principle simulate the dynamics of any closed $N$-qubit system.
That is, we can let our quantum computer mimic the time evolution
corresponding to any Hamiltonian we were given by some theorist
and then perform some measurements and check whether the results,
and with them the given Hamiltonian, really fit to the physical
system in the laboratory. Despite the naiveness of this description, one
crucial point here is whether or not the simulation can be
implemented efficiently on our quantum computer. In fact, it can,
as long as the Hamiltonian \begin{equation} \label{eqHsum}
H=\sum_l^L H_l
\end{equation} is again a sum of {\em local}
Hamiltonians $H_l$ acting only on a few particles\footnote{That
is, every $H_l$ involves at most a number of particles which is
independent of $N$.}. The basic idea leading to this result is the
following \cite{LloydSim}:

The evolution according to each $H_l$ can be easily simulated,
i.e., with an overhead which does not grow with $N$. Since the
different $H_l$ do in general not commute we have $\prod_l e^{-i
H_l t} \neq e^{-i H t}$. However, we can exploit Trotter's formula
\begin{equation} \label{eqTrotter}
\lim_{k\longrightarrow\infty}\left(\prod_{l=1}^L e^{-i H_l \frac{t}k}
 \right)^k=e^{-i H  t}
\end{equation}in order to move in the direction in Hilbert space
corresponding to $H$ by concatenating many infinitesimal moves
along $H_1, H_2,\ldots$. To use Lloyd's metaphor, this is like
parallel parking with a car that can only be driven forward and
backward. In fact, this is everyday life not only for car drivers
but also for people working in nuclear magnetic resonance where
sophisticated pulse sequences are used in order to drive a set of
spins to a desired state. The important point is, however,  that
in such a way $e^{-iHt}$ can be efficiently approximated with only
a polynomial number of operations. Moreover, the number of
required operations  scales as ${\cal O}(\mbox{poly}(1/\epsilon))$
with the maximal tolerated error $\epsilon$.

Evolutions of closed discrete systems are not the only things
which can be simulated efficiently. If for instance the terms
$H_l$ in Eq.\ (\ref{eqHsum}) are tensor products and $L$ is a
polynomial in $N$, this works as well \cite{Mike}. Moreover,
evolutions of open systems, approximations of systems involving
continuous variables \cite{BT97,Zal98}, systems of
indistinguishable particles, in particular fermionic systems
\cite{AL97}, and equilibration processes \cite{TD98} have
been studied.

Since the simulation of quantum systems becomes already an
interesting application for a few tens of qubits, we will see it
in the laboratories far before a ``Shor class'' quantum computer that
strikes classical factoring algorithms (and thus requires
thousands of qubits) will be build \cite{Jane}. In fact, we do not
even need a full quantum computer setup, i.e., the ability to
implement a universal set of gates, in order to simulate
interesting multi-partite quantum systems \cite{Bloch}.

\section{Quantum error correction}\label{SecQEC}

Quantum error correction aims at
protecting the coherence of quantum states
in a quantum computation against noise.
This noise is due to some physical interaction of
the quantum systems forming the quantum computer with
their environment, an interaction which can never be entirely avoided.
It turns out that reliable quantum computation is indeed
possible in the presence of noise, which was
one of the genuinely remarkable insights in this research
field. The general idea of quantum error correction is to
encode logical qubits into a number of physical qubits.
The whole quantum computation is hence
performed in a subspace of a larger dimensional Hilbert space,
called the \emph{error correcting code subspace}.
Any deviation from this subspace guides into an
orthogonal \emph{error subspace}, and can hence be detected
and corrected without losing the coherence
of the actual encoded states \cite{PreskillError}.
Quantum error correcting codes have the
ability to correct a certain finite-dimensional subspace
of error syndromes. These error syndromes could for example
correspond to a \emph{bit-flip error} on a single qubit.
Such bit-flip errors are, however,
by no means the only type of error that can occur to a single qubit.
In a \emph{phase flip error} the relative
phase of $|0\rangle$ and $ |1\rangle$ is interchanged.
Quantum error correcting codes can be constructed that
correct for such bit-flip and phase errors or both. In a quantum computing context,
this error correction capability is yet still not sufficient.
It is the beauty of the theory of quantum error correcting codes that
indeed, codes can be constructed that have the ability
to correct for a \emph{general error} on a single qubit (and
for even more general syndromes). What this means we
shall see after a first example.


\subsection{An introductory example}
The simplest possible encoding that protects at least against a very
restricted set of errors is the following:
Given a pure state of a single qubit with state vector $|\Psi\rangle
 =\alpha |0\rangle + \beta|1\rangle$.
This state can be protected against bit-flip errors of single qubits by means of the
\emph{repetition encoding}
$|0\rangle \mapsto |0,0,0\rangle$ and $|1\rangle \mapsto |1,1,1\rangle$,
such that $|\Psi\rangle$
is encoded as
\begin{eqnarray}
|\Psi  \rangle=\alpha |0\rangle + \beta|1\rangle  \mapsto \alpha |0,0,0\rangle+\beta   |1,1,1\rangle.
\end{eqnarray}
This encoding, the idea of which dates back to
work by Peres as early as 1985 \cite{Peres},
can be achieved by means of
two sequential CNOT gates to qubit systems initially
prepared in $|0\rangle$. Note that it does not amount to a
copying of the input state, which would be impossible anyway.
If an error occurs that manifests itself in a
\emph{single} bit-flip operation
to any of the three qubits, one can easily verify that one
obtains one out of
four mutually orthogonal states: These states correspond to no error at all,
and a single bit flip error to any of the three qubits.
This encoding, while not yet being a quantum error correcting code in the
actual sense,
already exemplifies an aspect of the theory: With a subsequent
measurement that indicates the kind of
error that has occurred, no information can be inferred
about the values of the coefficients $\alpha$ and $\beta$.  A
 measurement may hence enquire about the error
without learning about the data.

While already incorporating a key idea,
it is nevertheless not a particularly good encoding to protect against errors:
If a different error than a bit-flip occurs, then the measurement followed
by an error correction cannot recover the state. Moreover, and maybe more seriously,
the state cannot be disentangled from the environment, if the error is due to some
physical interaction entangling the state with its
environment. Let us consider the map involving
the qubit undergoing the error and the
environment, modeled as a system
starting with state vector $|\Psi_0\rangle$,
according to
\begin{equation}
|0,\Psi_0 \rangle \mapsto  |1,\Psi_0 \rangle,\,\,   \rm{ and } \,\, |1,\Psi_0\rangle
\mapsto  |0,\Psi_1
\rangle,
\end{equation}
such that the environment becomes correlated with the qubit undergoing the
error. This is a process typically referred to as decoherence.
The above encoding cannot correct for such an error
and recover the original state. Such an entangling error, however, correponds rather to
the generic situation happening in realistic errors.
In Preskill's words, the manifesto of quantum error correction is to
fight entanglement with entanglement \cite{PreskillError}.
What is meant is that
the unwanted but unavoidable entanglement of the system with its environment should
be avoided by means of skillfully entangling the systems in a quantum error
correcting code, followed by appropriate correction.
%
%

\subsection{Shor code}  There are, notably, error correcting codes that can
correct for any error inflicted on a single qubit of the code block. That such
quantum error correcting codes exist was first noted by Steane and Shor in
independent seminal
work in 1995 and 1996 \cite{Steane,ShorError}.
\emph{Shor's $9$ qubit code} is
related to the above repetition code
by encoding again
each of the qubits of the codewords into three other qubits, according
to $|0\rangle \mapsto (|0,0,0\rangle + |1,1,1\rangle )/\sqrt{2}$ and
$|1\rangle \mapsto (|0,0,0\rangle - |1,1,1\rangle )/\sqrt{2}$. If effect, in the
total encoding each logical qubit is encoded in the state of $9$ physical qubits, the
codewords being given by
\begin{eqnarray}
    |0\rangle \mapsto
    (|0,0,0\rangle + |1,1,1\rangle )
    (|0,0,0\rangle + |1,1,1\rangle )
    (|0,0,0\rangle + |1,1,1\rangle )/\sqrt{8},&&\\
     |1\rangle \mapsto
    (|0,0,0\rangle -|1,1,1\rangle )
    (|0,0,0\rangle - |1,1,1\rangle )
    (|0,0,0\rangle - |1,1,1\rangle )/\sqrt{8}.&&
\end{eqnarray}
In a sense, the additional encoding of the repetition code mends
the weaknesses of the repetition code itself. Such an encoding of the encoding
is called a \emph{concatenation of codes}, which plays an important role in quantum error correction.
What errors can it now correct? If the environment is initially
again in a pure state
associated with state vector $|\Psi_0\rangle$, then the most general
error model leads to the joint state vector
\begin{eqnarray}
    (\alpha |0\rangle + \beta |1\rangle) |\Psi_0\rangle &=&
    (\alpha |0\rangle + \beta |1\rangle) |\Psi_0\rangle+
    (\alpha |1\rangle + \beta |0\rangle) |\Psi_1\rangle\nonumber\\
    &+&
    (\alpha |0\rangle - \beta |1\rangle) |\Psi_2\rangle+
    (\alpha |1\rangle - \beta |0\rangle) |\Psi_3\rangle,
\end{eqnarray}
where no assumption is made concerning the state vectors $|\Psi_0\rangle$,
$|\Psi_1\rangle$, and
$|\Psi_2\rangle$, and
$|\Psi_3\rangle$.
One particular instance of this map is the one where
\begin{equation}
    |\Psi_0\rangle =|\Psi_2\rangle =|0\rangle, \,\,
    |\Psi_1\rangle =|1\rangle, \,\,
     |\Psi_3\rangle = -|1\rangle.
\end{equation}
One can convince oneself that when disregarding the state of the environment
(reflected by the partial trace), this error is a quite radical one: in effect, it is
as if the qubit is discarded right away and replaced by a new one,
prepared in $| 0\rangle$.
The key point now is that the Shor  code has the ability to correct for any
such error if applied to only one qubit of the codeword, and completely
disentangle the state again from the environment. This includes the complete
loss of a qubit as in the previous example.
In a sense, one might say
that the continuum of possible errors is discretized leading to orthogonal
error syndromes that can be reliably distinguished with  measurements,
and then reliably corrected for. But then, one might say, typical errors do not only affect
one qubit in such a strong manner, but rather, all qubits of the codeword are
exposed to errors. Even then, if the error is small and of the order ${\cal O}(\varepsilon)$
in $\varepsilon$ characterizing the fidelity of the affected state versus the input,
after error correction it can be shown to be of the order  ${\cal O}
(\varepsilon^2)$.

\subsection{Steane  code}
Steane's $7$ qubit  quantum error correcting code is a good example how
the techniques and the intuition from classical error correction can
serve as a guideline to construct good quantum error correcting codes
\cite{Steane,Steane2}.
It is closely related to a well-known classical code, the $[7,4,3]$-\emph{Hamming code}.
Starting point is the \emph{parity check matrix} of the $[7,4,3]$- Hamming code
given by
\begin{eqnarray}
    h=
    \left(
    \begin{array}{ccccccc}
    0 & 0& 0& 1& 1 &1& 1\\
    0 &1 &1 &0 &0 &1 &1 \\
    1 &0 &1 &0 &1 &0 &1\\
    \end{array}
    \right).
\end{eqnarray}
Codewords of the classical Hamming code are all binary words $v$ of length $7$
that satisfy $hv^T=0$, which is meant as addition in ${\mathbbm{Z}}_2$.
It is a straightforward exercise to verify that there are in total 16 legitimate
codewords (the kernel of $h$ is four-dimensional).
In the classical setting, if at most a single unknown
bit flip error occurs to a word $v$, leading to
the word $v'$,  it can be easily detected: if the error happens on the $i$-th
bit, then, from the very construction of $h$,
$hv'^T$  is nothing but
a binary representation of $i$, indicating the position of the error. If $hv'^T=0$,
one can conclude that $v'=v$, and no error has occurred.

\begin{figure}
\centerline{
\includegraphics[width=9.5cm]{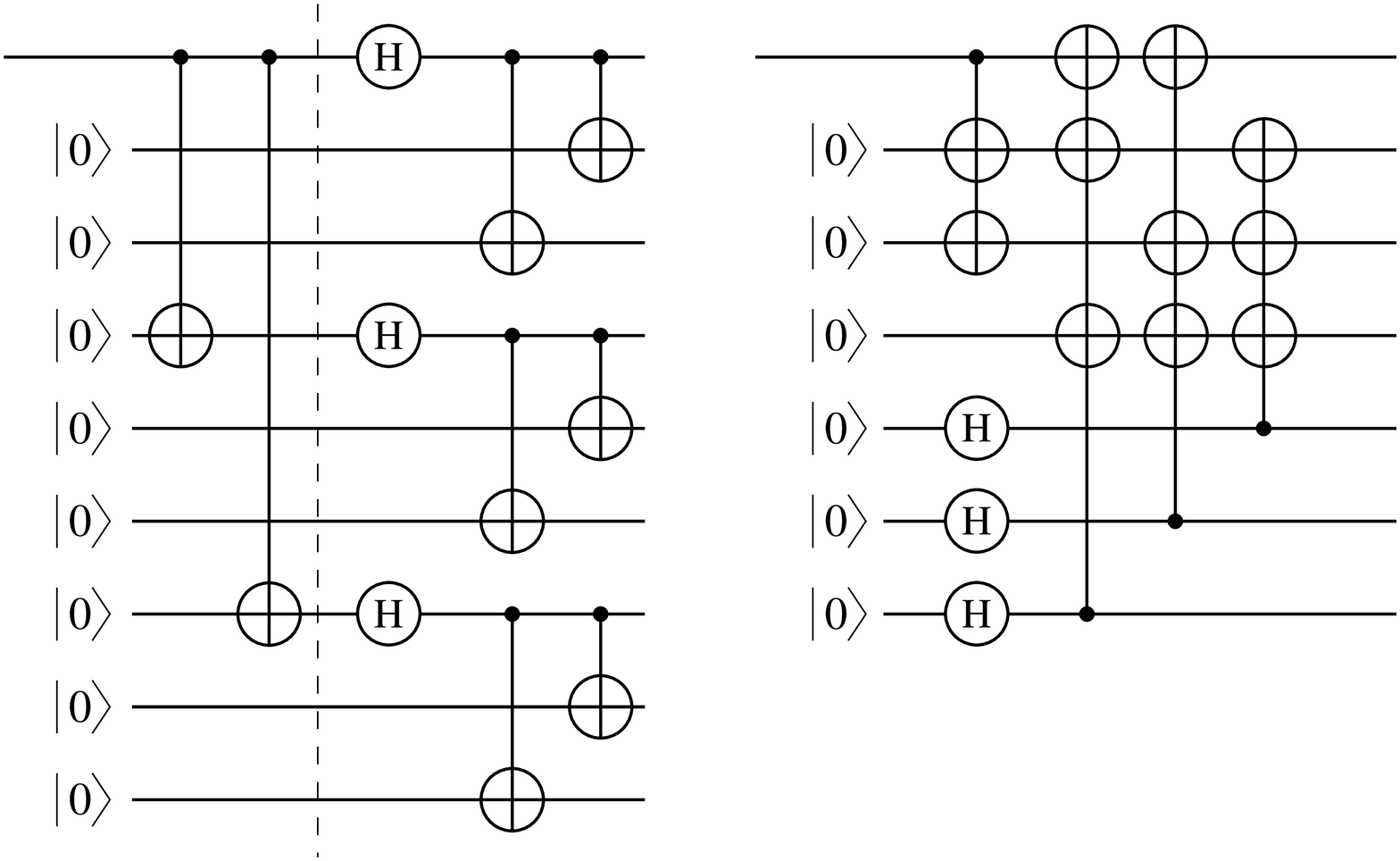}
}
    \caption{The encoding circuits of the Shor (left) and Steane (right) quantum
    codes. To the left of the dotted line, the depicted
    circuit corresponds to the repetition code. The first line corresponds
    to the input qubit.}\label{fig6}
\end{figure}

The \emph{$7$ qubit Steane code} draws from this observation.
It is now defined as follows: For the logical
$|0\rangle$, the quantum codeword is the superposition
of the eight codewords of the classical Hamming code with an odd number of $0$s,
represented in terms state vectors. The latter term means that the binary word
$x_1,...,x_7$ is represented as $| x_1,...,x_7\rangle$.
The logical $|1\rangle$ is encoded in a
similar state vector corresponding to the even number of $0$s. That is,
\begin{eqnarray}
    |0\rangle & \mapsto & ( |0,0,0,0,0,0,0\rangle + |0,0,0,1,1,1,1\rangle + |0,1,1,0,0,1,1,1\rangle\nonumber\\
    &+&
     |0,1,1,1,1,0,0\rangle + |1,0,1,0,1,0,1\rangle + |1,0,1,1,0,1,0\rangle\nonumber\\
    &+&
     |1,1,0,0,1,1,0\rangle + |1,1,0,1,0,0,1\rangle)/\sqrt{8} ,\\
     |1\rangle & \mapsto &
     (|0,0,1,0,1,1,0\rangle + |0,0,1,1,0,0,1\rangle + |0,1,0,0,1,0,1\rangle\nonumber\\
    &+&
     |0,1,0,1,0,1,0\rangle + |1,0,0,0,0,1,1\rangle + |1,0,0,1,1,0,0\rangle\nonumber\\
    &+&
     |1,1,1,0,0,0,0\rangle + |1,1,1,1,1,1,1\rangle/\sqrt{8} .
\end{eqnarray}
The central idea is now that  in the quantum situation one can make use of the idea
how the syndrome is computed in the classical case. When
appending a system consisting
of three qubits, the transformation
$|v'\rangle | 0,0,0\rangle \mapsto |v'\rangle| h v'\rangle$ can be realized in a unitary
manner, and the  measurement of the state of the additional qubits
reveals the syndrome. But this procedure,
one might be tempted to think, is merely
sufficient to correct for bit flip errors, from the construction of the $[7,4,3]$-Hamming code.
This is not so, however: a rotation of each qubit of the quantum codewords
with a Hadamard gate $H$
as described in section \ref{QvsC} with $|0\rangle
\mapsto( |0\rangle+ |1\rangle)/\sqrt{2}$ and
 $|1\rangle
\mapsto( |0\rangle- |1\rangle)/\sqrt{2}$
will yield again a superposition of binary
words. In fact, it is again a superposition of Hamming codeword, and bit flip errors in this
rotated basis correspond to phase flips in the original basis. So applying the
same method again will in fact detect all errors.
The encodings of the Shor and the Steane code are shown in Fig.\ \ref{fig6}.

\subsection{CSS and stabilizer codes} The formalism of
\emph{Calderbank-Shor-Steane codes} \cite{Calderbank,Steane2},
in short CSS codes,
takes the idea seriously that the theory of linear
codes can almost be translated into a theory of quantum error
correcting codes. Let us remind ourselves what
$[n,k,d]$ in the above notation specifying the classical
Hamming code stands for:
$n$ is the length of the code,
$k$ the dimension, and $d$ the distance,
the minimum Hamming distance between any two codewords.
At most $\lceil (d-1)/2 \rceil$ errors can be corrected by such a code.
Any such linear code is specified by its \emph{generator matrix} $G$, which
maps the input into its encoded correspondent. The parity check matrix $h$ can
be easily evaluated from this generator matrix. Associated with any linear
code is its \emph{dual code} with generator matrix $h^T$.
The construction of \emph{CSS codes} is based not on
one, but on two classical codes: on both a $[n_1, k_1, d_1]$ code $C_1$
and a  $[n_2, k_2, d_2]$ code $C_2$ with $C_2\subset C_1$,
such that both the former code and the dual of the latter
code can correct for $m$ errors. The quantum error correcting
code is then constructed for each codeword $x_1$ of $C_1$
as a superposition over codewords of $C_2$, again represented as
pure states of qubits.

With this construction, much of the power
of the formalism of classical linear error correcting codes can be
applied. It turns out that with such CSS codes, based on
the classical theory, up to $m$ errors can be detected and corrected,
indicating that good quantum error
correcting codes exist that can correct for more than general errors on single
qubits. The above Steane code is already an example for a CSS code,
but one that corrects only for a single error.
Is Steane's $7$ qubit quantum code the shortest quantum code
that can correct for a general error to a single qubit? The answer is no,
and it can be shown that five qubits are sufficient, as has first been
pointed out by Laflamme, Miquel, Paz and Zurek on the one hand
\cite{Laflamme},
and Bennett et al. \cite{BennettError}
on the other hand.
What can also be shown, in turn
is that no even shorter quantum code can exist with this capability.
This is an important insight when considering the hardware resources
necessary to design a quantum computer incorporating error correction.

This five qubit code is a particular instance of
a so-called \emph{stabilizer code} \cite{Gottesman}.
The stabilizer formalism is a very powerful formalism to
grasp a large class of unitary quantum operations on states, as
well as state changes under  measurements in the
computational basis. Essentially, instead of referring to the states
themselves, the idea is to specify the operators that ``stabilize the state'',
i.e., those operators the state vector is an eigenvector of
with eigenvalue 1.
It turns out that it is often by far easier and more transparent to specify
these operators than the state vectors. The power of the stabilizer formalism
becomes manifest when considering the \emph{Pauli group}, i.e., the
group of all products of the Pauli matrices and the identity with
appropriate phases. Based on this stabilizer formalism,
the important class of stabilizer codes can be constructed, which are
a genuine generalization of the CSS codes, and embodies also the
$9$ qubit Shor code. But the significance of the stabilizer formalism
goes much beyond the construction of good quantum error correcting codes.
The  \emph{Gottesman-Knill theorem} that has been mentioned previously in
section \ref{QvsC} can for example be proved using this formalism.

There is a notable link between stabilizer codes and quantum
error correcting codes based on graphs.
 A large class of quantum
error correcting codes
can be constructed based on a graph, where edges, roughly speaking, reflect
an interaction pattern between the quantum systems of the quantum codewords
\cite{Gottesman,Schlinge}.
It turns out that these \emph{graph codes} present an intuitive way of constructing
error correcting codes, and they exactly correspond to the stabilizer codes.
It is an interesting aspect that the \emph{graph states} \cite{Schlinge,Graph}
associated with graph codes can
serve also a very different purpose: they themselves form a universal resource
for measurement-based {\em one-way quantum computation} \cite{Oneway}.
In this scheme, a particular instance of a graph state is initially
prepared as a resource for the quantum computation.
Implementing a quantum algorithm amounts to performing
measurements on single qubits only (but not necessarily in the
computational basis), thereby realizing an effective unitary
transformation on the output qubits.

\subsection{Fault tolerant quantum computation}

Very nice, one might say at this point, it is impressive that
errors affecting quantum systems can be corrected for. But
is there not a crucial assumption hidden here? Clearly,
when merely storing quantum states, errors are potentially
harmful, and this danger can be very much attenuated indeed
by means of appropriate quantum error correction.
But so far, we assumed that the encoding and decoding of the quantum
states can be done in a perfectly reliable manner, without errors at all.
Given the degree of complexity of the circuits necessary to do such an
encoding (see, e.g., Fig.\ \ref{fig6}), amounting essentially to a quantum computation,
it does not seem very natural to assume that this computation can be done without
any errors. After all, one has to keep in mind that the whole procedure
of encoding and decoding complicates the actual computation,
and adds to the hardware requirements.

It was one of the very significant insights in the field that this assumption
is, unrealistic as it is, not necessary. In the recovery process, errors
may be allowed for, leading to \emph{fault tolerant recovery}, as has been shown
in seminal work by Shor \cite{ShorFaultTolerant},
with similar ideas
having been independently developed by Kitaev \cite{KitaevFaultTolerant}.
This is possible as long as the error rate in this process
is sufficiently low. But then, it might not be optimal to first encode, then later,
when appropriate, decode, perform a quantum gate, and then encode the state
again. Instead, it would be desirable to find ways of implementing a universal
set of gates in the space of the encoded qubits itself. This leads to the
theory of \emph{fault tolerant quantum computation}. That this is possible
has again been shown by Shor \cite{ShorFaultTolerant}, who devised
fault tolerant circuits
for two-qubit CNOT gates, rotations and three-qubit \emph{Toffoli gates} acting
as $|x,y,z\rangle\mapsto  |x,y,z\oplus x y\rangle$ \footnote{Note that
quite surprisingly, Toffoli and Hadamard gates alone are already
universal for quantum computation, abandoning the need for general
single qubit
%
rotations \cite{DoritHadamardToffoli,KitaevSim}.}.
This might still not be enough: from quantum error correction as described
before alone, it is not clear how to store quantum information to an
arbitrarily long time to a high fidelity. Knill and Laflamme demonstrated
that this is possible with \emph{concatenated encoding}, meaning
that the encoded words are encoded again to some degree of hierarchy,
and appropriate error detection and correction is performed
\cite{KnillLaflamme,KnillLaflammeZurek}.
Uniting these ingredients,
it became evident
%
that a threshold for the required accuracy for general
fault-tolerant quantum computation can be identified, allowing
in principle for arbitrarily long quantum computation with high
fidelity. Several inequivalent
{\em threshold theorems} asking essentially for only a constant error rate
have been developed that holds under a number of different assumptions
 \cite{PreskillError,BenOr,KitaevFaultTolerant,Gottesman,KnillLaflammeZurek}.
 Such schemes for achieving reliable quantum computation at a constant
 error rate can be achieved with a polylogarithmic overhead in both 
 time and space of the computation to be performed. 
 Hence, the additional cost in depth and  
size of the quantum circuit is such that the superiority of quantum algorithms
like Grover's and Shor's algorithms over their classical counterpart is
essentially preserved.


%
%

So in a nutshell, quantum error correction, together with techniques
from fault-tolerant quantum computation, significantly lessen the
threat posed by unavoidable decoherence processes
any quantum computer will suffer from. To preserve the coherence
of the involved quantum
over the whole quantum computation remains the central
challenge in a realization. The theory of quantum error correction, yet,
shows that the pessimism expressed in the mid 1990ies, culminating in the
statement that these
daunting problems cannot be overcome as a matter of principle, was
not quite appropriate.

\section{How to build a quantum computer?}\label{Secimp}

We have seen so far what purposes a quantum computer may
serve, what tasks it may perform well, better than any classical computer,
and have sketched what the underlying computational model is like. Also, ways have been
described to fight decoherence that is
due to the coupling to the environment, and eventually
to the same devices that are designed to perform the read-out.
The crucial
question remains: how can a quantum computer be built? What are
the physical requirements to appropriately isolate a quantum computer from its
environment? What is the physical hardware that can keep
the promise of the quantum computer as
a supercomputing device?

Needless to say, there is no satisfactory answer to these questions
so far. On the one hand, progress has been made in recent
years in the experimental
controlled manipulation of very small
quantum systems that can not be called other than spectacular, in a way
that was not imaginable not long ago. Quantum gates have been implemented
in the quantum optical context, and with nuclear magnetic resonance (NMR) techniques, even small
quantum algorithms have been realized.
On the other hand,
however, it seems fair to say that a universal quantum computer as a physical device
that deserves this name is still in the remote future. The only thing that seems safe
to say is that none of the current experimental efforts probably
deals with exactly the physical
system that will be used in an eventual realization of a Shor class quantum computer.
Supposably,
completely new ways of controlling individual quantum systems will
have to be devised, potentially combining previous ideas from quantum optics and solid
state physics.
Any such implementation will eventually have live up to some requirements that have
maybe most distinctly been formulated by DiVincenzo as generic requirements
in practical quantum computation \cite{Criteria},
see Fig.\ \ref{fig10}.
It is beyond the scope of this chapter to give an introduction to
the very rich literature on physical implementations of quantum computers. After all,
this is the core question that physicists seek to address in this field. Instead,
we will sketch a few key methods that have been proposed as potentially
promising methods or that have already been demonstrated in experiments.

\begin{figure}

\centerline{
\begin{minipage}{.9\textwidth}
\begin{tabular}{|ll|}
\hline\hline
(i) & Scalable physical system with well-characterized qubits\\
(ii) &  Ability to initialize the state of the qubits to a simple fiducial state\\
 (iii) & Long decoherence times, much longer than the gate operation time\\
 (iv) & Universal set of quantum gates\\
 (v) & Qubit specific measurement capability\\
 \hline\hline
\end{tabular}
\end{minipage}
}

\caption{The DiVincenzo criteria of what requirements must be met in
any physical implementation of a quantum computer.}\label{fig10}
\end{figure}

\subsection{Quantum optical methods}

Among the most
promising methods to date are quantum optical methods where
the physical qubits correspond to  \emph{cold ions in a linear trap}, interacting with
laser beams. A plethora of such proposals have been made, dating back to seminal
work by Cirac and Zoller \cite{CiracZoller}.
In the latter proposal, qubits are identified with
internal degrees of freedom of the ions, which are assumed to be two-level
systems for practical purposes. Single qubit operations can be accomplished
by means of a controlled interaction with laser light, shone onto the ions
by different laser beams that can individually address the ion. The ions repel
each other by Coulomb interaction, forming a string of ions with adjacent ions
being a couple of optical wavelengths apart from each other.
More challenging, of course,
is to find ways to let two arbitrary qubits interact to realize a two-qubit quantum gate.
This can be achieved by means of exciting the collective motion of the canonical degrees
of freedom of the ions with lasers, i.e., by using the
lowest level collective \emph{vibrational modes}.
Several refinements of this original proposal aim at
realizing the gates faster, and in a way that  does not
require extremely low temperatures or is less prone to decoherence
\cite{Milburn}.
Such quantum gates have already been realized in experiments,  notably
the implementation of two-qubit quantum gates
due to work by Monroe and co-workers with a single ion
\cite{Monroe} and Blatt and co-workers \cite{Blatt} with a two-ion
quantum processor.

Alternatively to using the motional degrees of freedom
to let quantum systems interact, this goal can be achieved by means of the tools
of \emph{cavity quantum electrodynamics} (cavity QED)
\cite{Pellizzari,Turchette}.
The key idea is to
store neutral atoms inside an optical cavity formed for example by two optical
super-mirrors.  The interactions required
to perform two-qubit gates are moderated by means of the interaction of the
atoms with a single quantized mode of a high-$Q$ optical cavity.
In Ref.\ \cite{Pellizzari}
%
%
it is assumed that adjacent atoms are separated by a few wavelengths of the cavity mode,
interacting with laser beams in an individual manner (standing qubits),
%
%
but also atomic beams passing through the
cavity have been considered, both theoretically and experimentally (flying qubits).
%
%
Two  regimes can in general be distinguished: the strong coupling limit, where
coherent atom-cavity dynamics            dominates cavity losses and spontaneous emission,
and the bad cavity limit, where
cavity loss rate is much larger     than the atom-cavity coupling.

Still a quantum optical setting, but without a quantum data bus in the closer
sense are proposals that make use of {\em controlled collisions} of cold atoms.
This can be realized for example with neutral atoms
in \emph{optical lattices}, where direct control over
single quantum systems can be achieved \cite{Controlled}.

Not to be confused with the classical
optical computer, in the \emph{linear 
optical quantum computer} 
the qubits are encoded in the state of 
field modes of light \cite{Optical}.
The state is manipulated by means of
optical elements such as beam splitters, mirrors, phase shifts, and
squeezers. The advantage -- that photons are not very prone to
decoherence -- is at the same time the disadvantage, as letting them interact is difficult as well as realizing strong Kerr 
non-linearities without significant losses.
Yet, in order to circumvene the latter problem, 
instead of requiring that a given task is accomplished
with unit probability, one may effectively realize the required
non-linear interactions by means of measurements of the photon 
number. This is possible at the expense that the scheme becomes probabilistic. 
Notably, Knill, Laflamme, and Milburn have proposed a 
near-deterministic scheme for 
universal quantum computation 
employing optical circuits that merely 
consist of passive linear optical elements (hence excluding squeezers)
together with photon counters that have the ability 
to distinguish $0$, $1$, and $2$ photons \cite{KLM}. 

Finally, the vibrational modes of molecules can be employed
to serve as qubits in {\em molecular quantum computers} \cite{Molecular}.
Both single qubit and two-qubit gate can be implemented
in principle by suitably shaped femtosecond laser pulses the form of which can be
computed by applying  techniques from control theory.
Drawbacks are
problems related to the scalability of the set-up.

\subsection{Solid state approaches}

As an alternative to
quantum optical settings serve solid state approaches.
Several different systems have been considered so far, including
proposals for
\emph{quantum dot} quantum computers with dipole dipole coupling.
Ideas from solid state physics and cavity QED can be combined
by considering solid-state quantum computers where gates
can be realized by controlled interaction between two distant
quantum dot spins mediated by the field of a high-$Q$ microcavity
\cite{Imamoglu,Imamoglu2}.
The Kane proposal is concerned with a
\emph{silicon-based nuclear spin quantum computer},
where the nuclear spins of donor atoms in doped silicon devices correspond to
the
physical qubits \cite{Kane}. The appeal of the proposal due to Ladd is that
it sketches a silicon quantum computer that could potentially
be manufactured using current fabrication techniques with
semiconductor technology and current measurement techniques \cite{Ladd}.
Finally, SQUIDs, \emph{superconducting quantum interference devices},
with the quantized flux serving as the qubit could be candidates
for a physical realization of a quantum computer.

\subsection{NMR quantum computing} Probably the most
progressed technology so far in a sense
is bulk ensemble quantum computation based on
\emph{nuclear magnetic resonance}
(NMR) techniques \cite{NMR1,NMR4}.
This idea is different from the previously described ones
in that it is not even attempted to control the state of individual quantum systems, trapped
or confined in an appropriate way. Instead, the state of
nuclear spins of $10^{20}-10^{23}$ identical molecules
is manipulated using the
well-developed tools from NMR technology.
Bulk techniques are used
not only for the reason that then the standard machinery of
NMR is available, but also because the nuclear
spin state
of a single molecule can hardly be properly prepared.
This set-up literally allows for quantum computation
with a cup of coffee.
Single qubit gates can fairly easily be realized. With
appropriate hand-tailored molecule synthesis and a sophisticated magnetic field
pulse sequence a $7$-qubit NMR quantum computer has been realized that
implements a shortened and simplified version of Shor's algorithm \cite{NMR4}.
However, quantum computation with bulk NMR techniques comes
with a caveat. Although most progress has so far been made in this area,
it has been convincingly argued that the scalability of these kinds of proposals
is limited by serious problems: notably, the signal is exponentially reduced
in the number of qubits by effective pure state preparation schemes
in an exponentiall manner in the number of qubits \cite{MerminArgument}.

\section{Present status and future perspective}

In the information age, where DVDs, wireless LAN, RSA encryption
and UMTS are the antiquated technologies of tomorrow, quantum
information theory aims to understand the old rules of quantum
mechanics from the new perspective of information theory and
computer science. In contrast to some earlier approaches to a
better \emph{understanding} of quantum mechanics, this one is a
very pragmatic one, leaving aside all metaphysical issues of
interpretation and coining former apparent paradoxes into future
applications. The most challenging and outstanding of these
 is the universal quantum computer. Its potential is not yet fully
 understood. At the moment there are essentially two classes of very
 promising quantum algorithms: search algorithms based on Grover's
 database search and applications of the quantum Fourier transform
 like Shor's factoring and discrete logarithm algorithms\footnote{More recent
 progresses in this direction are polynomial-time quantum algorithms for estimating 
 Gauss sums \cite{vanDam} and solving Pell's equation \cite{Hallgren}.}.
 In  particular, the latter yield an exponential speed-up compared to
 the best known classical algorithms. For which other problems
 can we expect such a speed-up? The killer application would of
 course be a polynomial-time algorithm for NP-complete problems.
 Being optimistic one could consider results in adiabatic
 computing as supporting evidence for this desire. However, the
 optimality of the quadratic speed-up in search algorithms might
 be evidence to the contrary. Moderating our optimism a bit, we
 could try to find efficient quantum algorithms for problems which
 are believed to be hard classically but not NP-complete --- the
 hottest candidate of such problems is probably the graph isomorphism
 problem, for which despite considerable effort no efficient
 quantum algorithm has been found so far.

 What role does entanglement play in quantum computers? This
 question is in general not entirely answered yet. However, if we
 consider a quantum computer unitarily acting on a pure input
 state, then an exponential speed-up compared to classical
 computers can only be achieved if the entanglement present in
 intermediate states of the computation increases with size of the
 input \cite{JozsaEnt,VidalEnt}\footnote{As shown by Vidal \cite{VidalEnt} the evolution of a pure states of $N$ qubits
 can be simulated on a classical computer by using resources that grow
 linearly in $N$ and exponentially in the entanglement. Similarly
 the evolution of mixed states, on which the amount of correlations is restricted,
 can be efficiently simulated.
 Note that sub-exponential speed-ups like in Grover's search algorithm could also be achieved
 without entanglement or a restricted amount of it \cite{LloydSearch}.}. It appears that computations based
 on such (rather typical) quantum evolutions can in general not be simulated efficiently on
 classical computers.

 Let us finally speculate how a quantum computer will eventually
 look like. What will be its hardware? In the past, the most successful
 realization was NMR, where even small quantum circuits have been
 implemented. Unfortunately, it has been convincingly argued that this
 implementation is not scalable to larger circuits. For the near
 future ion traps and, in particular regarding the simulation of quantum systems, optical lattices seem to be quite
 promising, whereas in the remote future solid state realizations
 would be desirable.
 However, progress is never smooth:

\begin{center} \parbox{7.8cm}{\emph{``Where a calculator on the ENIAC is
equipped with 18,000 vacuum tubes and weighs 30 tons, computers in
the future may have only 1,000 tubes and perhaps only weigh 1 1/2
tons.'' (Popular Mechanics, March
 1949)}}\end{center}

\end{document}